\documentclass{article}
\usepackage{graphicx}
\usepackage{amsmath}

\input{tcilatex}

\begin{document}

\title{Slowdown of nonequilibrium dynamics in gapped `qubit' chains}
\author{D. Tygel, J. G. Carvalho, and G. G. Cabrera \\
Instituto de F\'{i}sica ``Gleb Wataghin'',\\
Universidade Estadual de Campinas (UNICAMP),\\
C. P. 6165, Campinas 13083-970 SP, Brazil}
\maketitle

\begin{abstract}
We solve the nonequilibrium dynamics of qubits or quantum spin chains
(s=1/2) modeled by an anisotropic XY Hamiltonian, when the initial condition
is prepared as a spatially inhomogeneous state of the magnetization.
Infinite systems are studied analytically, yielding a universal relaxation
driven by quantum fluctuations. Particular long-lived excitations are found,
for which the relaxation time diverges as a consequence of constructive
quantum interference at degenerate stationary points. Those states are
intrinsically entangled and may be of interest for performing quantum
computation. We also numerically analyze finite samples to assess the extent
of size effects.
\end{abstract}

\section{Introduction}

\noindent Magnetic quantum systems which displays a slow relaxation in the
nonequilibrium dynamics are good candidates for magnetic data storage
devices. Low dimensional magnetic systems, including magnetic chains and
magnetic molecules in molecule-based magnets, may allow for a substantial
reduction of the size of memory units to perform classical computation \cite
{mrs}. In addition, systems that develop quantum coherence in the nanometer
scale may be of interest for the fabrication of quantum computers. Quantum
computation is believed, at least in principle, to be much faster than
classical computation, and its accomplishment rests in the ability of a
quantum system to be in a superposition of many quantum states (\emph{%
entanglement}). This property is the basis of the so called \emph{quantum
parallelism}, which presents a lot of advantages in relation to its
classical counterpart, the one that is realized in a classical computer by
connecting a large number of processors for parallel operation. Still, there
is another important requirement for doing efficient quantum computation, a
condition that is related to quantum programming or quantum software: we
need robust Bell-like states, or stable entangled states devised through
constructive interference from the many quantum states available. The
coherence time of those states must be much larger than the time employed in
1-qubit or 2-qubit operations \cite{galindo}.

The present contribution is related to this latter issue. It has been shown
that quantum chains which are modeled through the Heisenberg Hamiltonian (or
variants such as the quantum $XY$ model), are intrinsically entangled \cite
{entangle}. The quantum $XY$ model can be practically realized in
low-capacitance Josephson junction arrays \cite{makhlin}; it also appears as
an effective Hamiltonian for the interaction of quantum dot spins \cite{dots}%
, and it is also relevant for other condensed matter and QED systems which
have been proposed as quantum computers \cite{galindo}. One important role
of the $XY$ model in quantum computation is that it can be used to construct
a \emph{swap} gate, which means that the time evolution can be consider as a
source of entanglement \cite{entangle}. Moreover, an interesting connection
between quantum phase transitions and entanglement has been recently
proposed, suggesting that entanglement plays an important role in quantum
critical phenomena \cite{entangle2}. Concerning the anisotropic $XY$ model,
a quantum phase transition is found as a function of the anisotropy
parameter and transverse field. The latter may be envisioned as a quantum
metal-insulator transition with the formation of a gap for the insulating
phase.

Here, we are able to construct analytic solutions of the anisotropic $XY$
model in the thermodynamic limit, which display a very slow relaxation after
being prepared in a non-homogeneous initial state. This is a peculiar
nonequilibrium problem, since our quantum system is closed (does not
exchange energy with the environment except when preparing the initial
condition), and its relaxation to the homogeneous state is exclusively
driven by quantum fluctuations and interference effects. Nonequilibrium
properties of quantum systems have been recently investigated by several
authors following different approaches \cite{antal,karevski,igloi}, but a
complete theory is still lacking. To test our solutions, we also study
finite systems through numerical calculations. Finite systems are
interesting for practical applications, like the ones mentioned above. An
important question to answer is whether a finite system will show similar
relaxation properties, at least during finite periods of time, and how to
monitor the magnitude of such periods.

In spite that we solve a well defined model, some properties of the
phenomenon seem to be universal for one dimensional systems. \emph{The key
point addressed in this paper is how to prepare special long-lived
superposition states in quantum magnetic chains, whose coherence is not
wiped out by the strong destructive interference that appear at very long
time}. We call those states as \emph{critical}, for reasons that will be
developed in the next sections.

A general nonhomogeneous initial state in a quantum spin chain can be
Fourier decomposed. We then study the nonequilibrium dynamics through the
time evolution of the Fourier component of the magnetization $\ <\mathbf{%
\vec{S}}_{Q}>_{t}$. This approach has several advantages in spite that $%
\mathbf{\vec{S}}_{Q}$ is not an observable: one can probe the relaxation
properties as a function of the wave vector $Q$, which is an important
parameter in the theory, and for the particular case of the $XY$ model (and
variants), the relaxation of the longitudinal component $<S_{Q}^{z}>_{t}$ is
independent of the initial state, except for a complex scale factor, \emph{%
i.e.} 
\begin{equation}
<S_{Q}^{z}>_{t}=<S_{Q}^{z}>_{0}\ f^{z}(Q,t)\ ,  \label{scale}
\end{equation}
where $f^{z}$ is a real universal function that depends on parameters of the
Hamiltonian. We are interested in solving the anisotropic $XY$ magnetic
chain with spin $1/2$ in the presence of a transverse magnetic field 
\begin{equation}
\mathcal{H}=2\hbar \omega _{J}\sum_{j=1}^{N}\ \left\{ rS_{j}^{z}+\left[
\left( 1+\gamma \right) S_{j}^{x}S_{j+1}^{x}+\left( 1-\gamma \right)
S_{j}^{y}S_{j+1}^{y}\right] \right\} ,  \label{hamiltonian}
\end{equation}
where $\hbar \omega _{J}$ is the exchange integral, $r=\left( \omega
_{0}/\omega _{J}\right) $ is the external transverse field in units of the
exchange constant, $0\leq \gamma \leq 1$ is the anisotropy parameter, $N$ is
the total number of spins in the chain, and $S_{j}^{\alpha }=\left(
1/2\right) \sigma _{j}^{\alpha }$ are spin 1/2 operators. Special cases of
this model are:

\begin{enumerate}
\item[i)]  the isotropic $XY$ model with a transverse field ($\gamma =0$);

\item[ii)]  the Ising model with a transverse field ($\gamma =1$);

\item[iii)]  no magnetic field ($r=0$).
\end{enumerate}

The chain is prepared in an arbitrary nonhomogeneous initial state and its
magnetization dynamics is calculated in exact form under very general
assumptions. In the thermodynamic limit ($N\rightarrow \infty $), and for
long times, the system displays irreversibility in the form of a power law
relaxation. This irreversibility can be ascribed to quantum interference
effects which encompass the system at asymptotically long times. In analogy
with classical optics, we handle this phenomenon using the stationary phase
method \cite{stationary}. As a general rule, most of the components
interfere destructively (shadow regions), and the relaxation is dominated by
the contribution of stationary points, where the interference is
constructive (illuminated zones). It may happen that some stationary points
become degenerate in the parameter space. In optics, we get surfaces where
the intensity of light diverges (\emph{caustics}). In our problem, we get
critical states whose relaxation time asymptotically diverges, yielding a
very slow relaxation. Those critical modes do not occur in all cases, and we
present a summary of our findings below.

Models i) and ii) were discussed in Ref. \cite{berim}. Also Berim and
Cabrera \cite{BC} and Berim, Berim and Cabrera \cite{BBC} have solved
similar problems for several versions of the 1-dimensional $XY$ model, which
include dimerization, bond alternation, and staggered magnetic field. All
the models, including the one presented here, are amenable of exact analytic
treatment by fermionization methods involving generalized Jordan-Wigner
transformations \cite{BBC,lieb}. \emph{Concerning relaxation properties, all
the models studied can be grouped into two families, whether the energy
spectrum has a gap or not }\cite{BBC}. Only gapped models develop critical
modes with anomalously slow relaxation properties. It remains the question
whether the above phenomenon is associated to a conservation law \cite{BBC}.
It is then interesting to study the behavior of Hamiltonian (\ref
{hamiltonian}), since the anisotropy in the $XY$ exchange and the transverse
magnetic field lift the spin rotational symmetry and the time reversal
symmetry, respectively. Moreover, one can monitor the spectral properties
(gap or not) with $\gamma $ and $r$. The present study confirms the general
prescription stated in Ref. \cite{BBC} for the $XY$ family, \emph{i.e.}
anomalous long-time tails in the relaxation are features of gapped
one-dimensional models. In this contribution, we will show that this
anomalous slowing down of the relaxation is due to both, constructive
quantum interference at degenerate or \emph{critical }stationary points and 
\emph{nesting }properties of the spectrum.

\section{The calculation}

Our calculation follows the general trends described in detail in Ref. \cite
{BC,BBC,condmat}. Hamiltonian (\ref{hamiltonian}), with periodic boundary
conditions, is written in terms of fermion variables $(c_{j},c_{j}^{\dagger
})$ using the Jordan-Wigner transformation: 
\begin{equation}
S_{j}^{x}=L_{j}(c_{j}+c_{j}^{\dagger })/2,\qquad
S_{j}^{y}=L_{j}(c_{j}-c_{j}^{\dagger })/2i,\qquad S_{j}^{z}=c_{j}^{\dagger
}c_{j}-1/2,  \label{JW}
\end{equation}
with $L_{j}$ being the sign string that adjusts the anticommutation
relations for the $(c_{j},c_{j}^{\dagger })$%
\begin{equation}
L_{j}=\prod_{l=1}^{j-1}\left( 1-2c_{j}^{\dagger }c_{j}\right) ,\qquad
L_{1}\equiv 1,\qquad L_{j}^{2}=1.  \label{JW1}
\end{equation}
For the fermion Hamiltonian ($c-$Hamiltonian), we solve the \emph{c-cyclic}
problem, where one neglects a correction term that comes from the boundary,
as it is usually done when studying the thermodynamic limit ($N\rightarrow
\infty $) \cite{lieb}. The \emph{c-cyclic }Hamiltonian is diagonalized in
two steps:

\begin{enumerate}
\item[i)]  we take the Fourier transform 
\[
b_{k}=\sqrt{\frac{1}{N}}\sum_{j=1}^{N}\ c_{j}\exp \left( i\ kj\right) , 
\]
with $k=2m\pi /N\ (-N/2\leq m\leq N/2-1)$. At this stage, the Hamiltonian is
not diagonal since the anisotropy couples the modes $\left( k,-k\right) $;

\item[ii)]  complete diagonalization is achieved with a Bogoljubov
transformation 
\begin{equation}
\eta _{k}=u_{k}b_{k}+iv_{k}b_{-k}^{\dagger }\ ,\qquad \eta _{k}^{\dagger
}=u_{k}b_{k}^{\dagger }-iv_{k}b_{-k}\ ,  \label{bogoljubov}
\end{equation}
with real coefficients $\left( u_{k},v_{k}\right) $ and Hamiltonian 
\begin{equation}
\mathcal{H}=\sum_{k}\ \hbar \Lambda _{k}\eta _{k}^{\dagger }\eta _{k}+C\ ,
\label{diagonal}
\end{equation}
where $C$ is a constant term and the quasi-particle dispersion relation is
given by 
\begin{equation}
\Lambda _{k}=2\omega _{J}\sqrt{\left( \gamma \sin \ k\right) ^{2}+\left(
r+\cos \ k\right) ^{2}}\ .  \label{energy}
\end{equation}
In the particle-hole representation, $\Lambda _{k}\geq 0$ and the ground
state is the vacuum for the $\eta $ operators. We note that the spectrum
given by (\ref{energy}) may present a gap as a function of the anisotropy $%
\gamma $ and transverse field $r$. Several examples are available:

\begin{enumerate}
\item  for $\gamma =1$ (Ising with transverse field), the spectrum presents
a gap at $k=\pi $ of magnitude 
\begin{equation}
\Delta =2\left| 1-r\right|  \label{gap1}
\end{equation}
in units of the exchange constant, which shows that there is a critical
value $r_{c}=1$ for the transverse field;

\item  when $r=0$ (no magnetic field), we get a gap at $k=\pi /2$ whose
value is 
\begin{equation}
\Delta =2\gamma \quad ,  \label{gap2}
\end{equation}
the isotropic $XY$ model being gapless;

\item  with magnetic field and anisotropy, when $r\leq 1-\gamma ^{2}$, the
gap is positioned at $k=\arccos \left( -\frac{r}{1-\gamma ^{2}}\right) $,
with value 
\begin{equation}
\Delta =2\gamma \sqrt{1-\frac{r^{2}}{1-\gamma ^{2}}}\ .  \label{gap3}
\end{equation}
Otherwise, its position and value are given by (\ref{gap1}).
\end{enumerate}
\end{enumerate}

Once the ground state and excitations have been determined, we proceed to
calculate the dynamical properties of the magnetization. This is done using
the identity \qquad 
\begin{equation}
\left\langle A\right\rangle _{t}=\mathrm{Tr}\left[ \rho (t)A\right] =\mathrm{%
Tr}\left[ \rho (0)A(t)\right] \ ,  \label{average}
\end{equation}
where $A(t)$ is an operator in the Heisenberg picture. This way, (\ref
{average}) relates the temporal evolution of the average at any time with
the initial state $\rho (0)$. We will probe the dynamics of the longitudinal
(parallel to the field) Fourier component of the magnetization 
\begin{equation}
\left\langle S_{Q}^{z}\right\rangle _{t}=\mathrm{Tr}\left[ \rho
(t)S_{Q}^{z}\right] =\mathrm{Tr}\left[ \rho (0)S_{Q}^{z}(t)\right] \ ,
\label{OP}
\end{equation}
with 
\[
S_{Q}^{z}\equiv \sum_{j=1}^{N}\ \ S_{j}^{z}\exp \left( i\ Qj\right) \ . 
\]
Exact closed forms for (\ref{OP}) are obtained when one assumes that the
initial density matrix is a functional of only one spatial component of the
spins, a situation that can be achieved in practice using strong
non-homogeneous magnetic fields along a given direction, to prepare the
initial state \cite{BC,BBC}. We will denote those possibilities as $\rho
(0,S^{\mu })$, with $\mu =x,y,z$ labeling the possible field directions. The
calculation follows the general trends described in Ref. \cite{BBC,condmat}.
The details will be given elsewhere \cite{else}. Note that the calculation
of the dynamics of the transverse components $S_{Q}^{x}$ and $S_{Q}^{y}$ is
much more involved, and it is not clear that one can obtain closed analytic
formulae. For the above components we only obtained numeric results in small
systems.

For the longitudinal component, we summarize the main steps of the
calculation below:

\begin{enumerate}
\item[\emph{(a)}]  transform $S_{Q}^{z}$ to fermion operators to get its
time dependence. This step leads to: 
\begin{eqnarray}
S_{Q}^{z}\left( t\right) &=&-\frac{N}{2}\delta _{Q,0}+\sum_{k=-\pi }^{\pi
-2\pi /N}\ \left\{ \left( u_{p}u_{q}e^{i\Omega ^{-}t}\ \eta _{p}^{\dagger
}\eta _{q}+v_{p}v_{q}e^{-i\Omega ^{-}t}\ \eta _{-p}\eta _{-q}^{\dagger
}\right) +\right.  \nonumber \\
&&  \label{mag} \\
&&\qquad \left. +\left( -iu_{p}v_{q}e^{i\Omega ^{+}t}\ \eta _{p}^{\dagger
}\eta _{-q}^{\dagger }+iv_{p}u_{q}e^{-i\Omega ^{+}t}\ \eta _{-p}^{\dagger
}\eta _{q}^{\dagger }\right) \right\} \ ,  \nonumber
\end{eqnarray}
where $p=k-Q/2$ , $q=k+Q/2$, $\Omega ^{\pm }(k,Q)\equiv \Lambda _{p}\pm
\Lambda _{q}$ , and $(u,v)$ are the coefficients in the transformation (\ref
{bogoljubov}). Note that in (\ref{mag}) there are contributions from
processes that do not conserve the number of particles;

\item[\emph{(b)}]  average with the initial density matrix. This means that
we have to calculate the averages $\left\langle \eta _{p}^{\dagger }\eta
_{q}\right\rangle _{0},\ \left\langle \eta _{-p}\eta _{-q}^{\dagger
}\right\rangle _{0},\ \left\langle \eta _{p}^{\dagger }\eta _{-q}^{\dagger
}\right\rangle _{0},$ and $\left\langle \eta _{-p}^{\dagger }\eta
_{q}^{\dagger }\right\rangle _{0}$ in (\ref{mag}) over the initial state;

\item[\emph{(c)}]  take the thermodynamic limit $N\rightarrow \infty $, 
\emph{i.e.} introduce an infinite number of degrees of freedom. As a
consequence, summations over the $k$-space are replaced by integrals over
the Brillouin zone. Using the symmetry of $k$, one can reduce the
integration to the interval $\left[ 0,\pi \right] $;

\item[\emph{(d)}]  find asymptotic analytic expressions for very long times, 
$t\rightarrow \infty $.
\end{enumerate}

Note the order of the limiting processes of the two last steps. Indeed, for
finite $N$, the limit $t\rightarrow \infty $ does not exist. We will comment
on this further in the following. For the intended scope of this paper, we
will just consider the behavior of $<S_{Q}^{z}>_{t}$ for the initial
ensemble $\rho (0,S^{z})$. All the cases, along with complete formulae, are
given in \cite{else}. We quote the exact result which comes from taking the
continuous limit in (\ref{mag}): 
\begin{eqnarray}
\left\langle S_{Q}^{z}\right\rangle _{t} &=&\frac{1}{2\pi }\left\langle
S_{Q}^{z}\right\rangle _{0}\int_{0}^{\pi }\ dk\ \left\{ h(k,Q)\ \cos [\Omega
^{-}(k,Q)\ t]+\right.  \label{SZ} \\
&&\quad \left. +g(k,Q)\ \cos [\Omega ^{+}(k,Q)\ t]\right\} \quad ,  \nonumber
\end{eqnarray}
where the frequencies $\Omega ^{\pm }$ are defined above and the functions $%
(h,g)$ are given by: 
\begin{equation}
\begin{array}{l}
h(k,Q)\equiv 1+\dfrac{\left( 2\omega _{J}\right) ^{2}}{\Lambda _{p}\Lambda
_{q}\ }[r^{2}+2r\cos k\ \cos Q/2+\cos ^{2}k\ (\cos ^{2}Q/2+ \\ 
\qquad \qquad \qquad +\gamma ^{2}\sin ^{2}Q/2)-\sin ^{2}k\ \left( \sin
^{2}Q/2+\gamma ^{2}\cos ^{2}Q/2\right) ]\ , \\ 
g(k,Q)\equiv 2-h(k,Q)\ .
\end{array}
\label{details}
\end{equation}
The frequencies $\Omega ^{\pm }$ which appear in (\ref{mag}) and (\ref{SZ}),
correspond to particle-particle, hole-hole and particle-hole excitation
processes that contribute to the time evolution of $S_{Q}^{z}$. Note that
for the regime $t\rightarrow \infty $, the integral (\ref{SZ}) is strongly
oscillatory, and integration over ordinary points leads to cancellations
effects (destructive interference). The asymptotic behavior, obtained
through the \emph{stationary phase method }\cite{stationary}, is dominated
by the contributions of stationary points, where we get constructive
interference. The detailed study of stationary points of $\Omega ^{\pm }$
depends on $\gamma ,\ r$ and $Q$. The points $k=0$ and $k=\pm \pi $ are
always solutions. In addition, the other stationary points are given as
roots of a polynomial of fifth degree of the variable $y=\cos k$. This means
that the general problem has no analytic solutions, unless a fortuitous
factorization is achieved. This way, we have only obtained analytic
expressions for two important particular cases: model ii), the Ising model
with transverse field ($\gamma =1$ and arbitrary $r$); and model iii), no
transverse field ($r=0$) and arbitrary anisotropy $\gamma $. As a general
rule, the contribution of stationary points leads to an asymptotic
relaxation with time in the form of a power law. We write 
\begin{equation}
\left\langle S_{Q}^{z}\right\rangle _{t}\sim
\sum_{n}J_{n}^{+}(t)+\sum_{n}J_{n}^{-}(t)\ ,  \label{asymptotic}
\end{equation}
where $(n,\pm )$ labels the stationary point of $\Omega ^{\pm }$
respectively. The number of stationary points and their degeneracy depend
only on the parameters of the Hamiltonian $(\gamma ,r)$ and the wave vector $%
Q$. The dominant time dependence of the $J_{n}^{\pm }(t)$ functions in (\ref
{asymptotic}) is given by 
\begin{equation}
J_{n}^{\pm }(t)=K_{n}^{\pm }\ \exp \left( i\Theta _{n}^{\pm }\ t\right) \
\left( \frac{t}{\tau _{n}^{\pm }}\right) ^{-\nu _{n}^{\pm }}\ ,
\label{powerlaw}
\end{equation}
using the same notation as above. The important quantities in (\ref{powerlaw}%
) are the relaxation rate $\tau ^{\pm }$ and the exponent $\nu ^{\pm }$,
which together determine the relative speed of the relaxation process. For
the cases we calculated analytically, they are functions of $(\gamma ,Q)$ or 
$(r,Q)$. In Fig.1, we display the relaxation times as function of $\gamma $
for a given value of the wave vector $Q$ (with $r=0$). The asymptotic
behavior is dominated by the contribution of five stationary points of $%
\Omega ^{+}$: two of them are related to $\tau _{1}$ (the points $k=0,\pi $%
), a third one ($k=\pi /2$) is associated with $\tau _{2}$, and from the two
remaining we get the same relaxation time $\tau _{3}$. The times $\tau _{2}$
and $\tau _{3}$ become asymptotically large in the vicinity of a critical
value of $\gamma $, which is indicated in the figure, where the
corresponding stationary points become degenerate. Some general comments are
in order:

\begin{enumerate}
\item[\emph{i)}]  the asymptotic behavior oscillates with a small number of
frequencies $\Theta _{n}^{\pm }$, which are functions of $(\gamma ,Q)$ or $%
(r,Q);$

\item[\emph{ii)}]  we get a whole set of exponents $\nu _{n}^{\pm }$ coming
from the different stationary points, but the relaxation for $t\rightarrow
\infty $ is driven by the smallest exponent $\nu _{0}$;

\item[\emph{iii)}]  depending on their variation in the parameter space,
some stationary points may become degenerate. This fact drastically affects
the relaxation. As remarked before, the $\tau $'s asymptotically diverge in
the neighborhood of critical points. Exactly at the critical values, one has
to go some steps further in the asymptotic expansion (as many as the order
of the degenerate point), yielding a discontinuous change of the $\nu _{0}$
exponent. In the example of Fig. 1, the exponent $\nu _{0}$ changes from $%
1/2 $ to $1/4$ at the critical $\gamma _{C}$ , signaling a slowing down of
the relaxation process;

\item[\emph{iv)}]  the above instance occurs at the loci of the \emph{%
critical }curves $Q=Q_{C}$ in the parameter space, with $Q_{C}$ given by: 
\begin{equation}
Q_{C}(\gamma =1,r)=\left\{ 
\begin{array}{l}
2\ \arccos r\ ,\quad r<1, \\ 
2\ \arccos \left( \frac{1}{r}\right) ,\quad r>1
\end{array}
\right. \ ,  \label{qcr}
\end{equation}
for the Ising case with transverse field, and 
\begin{equation}
Q_{C}(\gamma ,r=0)=\arccos \left( \frac{1-\gamma }{1+\gamma }\right) \
,\qquad \mathrm{for\ }0\leq \gamma <1,  \label{qcgamma}
\end{equation}
for the anisotropic case and no field. Note that the gapless cases $(\gamma
=1,r=1)$ and $(\gamma =0,r=0)$ yield $Q_{C}\rightarrow 0$, which is an
ordinary stationary point. Also note that the pure Ising case $(\gamma
=1,r=0)$ has no dynamics and has to be treated separately.
\end{enumerate}

The above results are exact in infinite chains, where the thermodynamic
limit implies a continuous spectrum. In contrast, finite systems present
quantum recurrences, with partial or total reconstruction of the initial
state. For very small samples, we are able to identify a number of Rabi-like
periods associated with the revival of the magnetization. Numerical
computations show that those periods or quasi-periods become longer with
increasing sizes, scaling almost linearly with $N$. This poses the question
of how to define the time interval during which the finite system
approximate the infinite system in a regular way, in the spirit of
finite-size scaling methods. To estimate the relaxation, we have adopted the
following procedure: \emph{a) }we numerically analyze the dynamics of finite
systems until the first partial reconstruction (this time depends on size
and $\gamma $), and; \emph{b)} we fit the time evolution to a power law
similar to the one given by (\ref{powerlaw}). This is not an easy task,
since there are several frequencies superimposed. A typical behavior of the
magnetization is shown in Fig. 2, for $N=100$ and for the same $Q$ wave
vector of Fig.1. We estimate the exponent from the envelopes of local maxima
and minima, which are shown in the figure. In Fig. 3, we display a summary
of our findings for $N=100$ spins. One observes a precursory slowing down of
the relaxation around the critical value predicted for the infinite chain
(see Fig. 1), with a typical rounding due to size effects. We comment those
results further in the next section.

\section{Conclusions}

One interesting feature of the anisotropic $XY$ model is the absence of any
spin conservation law (we exclude the pure Ising limit without transverse
field). This fact makes its dynamics \ `richer' than the one associated with
other models, where the total spin or one of the total spin projections are
conserved \cite{BC,BBC}. This appears as a source of entanglement, which is
relevant for applications in quantum computation. Another peculiarity of the 
$XY$ model is exhibited in (\ref{scale}) and (\ref{SZ}), meaning that the
dynamics of the longitudinal Fourier component of the magnetization is
independent of the initial state, which enters just through a scale factor
in the form of an average over the initial condition. This property is not
shared by other models, like the Heisenberg $XXZ$ model, where we found a
strong dependence of the relaxation process on the initial state \cite{xxz}.

From our analytic and numeric computations, we were able to identify
long-lived collective excitations, which have the character of dynamic spin
density waves with incommensurate wave vector $Q_{C}$. In the vicinity of $%
Q_{C}$, we get $\tau (Q)\rightarrow \infty $ asymptotically, thus reducing
the damping of those excitations in relation to a general wave vector $Q$.
Exactly at the point $Q=Q_{C}$, the exponent $\nu _{0}$ changes
discontinuously $(1/2\rightarrow 1/4)$ and the relaxation time $\tau $ has a
finite value. Note that \emph{critical }modes only appear when the spectrum
has a gap, in agreement with findings in other models of the $XY$ family 
\cite{berim,BC,BBC}.

Our theoretical results may be physically realized in a number of systems,
from which the most promising to perform quantum computation are Josephson
tunneling junction arrays. Quantum bits (`qubits') stored in low-capacitance
Josephson junctions (JJ) simulate mesoscopic spins, and the coupling between
JJ units mimic the spin-spin interactions that are usually found in several
spin models \cite{makhlin}; in particular one can obtain an effective
anisotropic $XY$ coupling with a gap in the spectrum. Nonhomogeneous initial
states considered in this paper, could be practically achieved in JJ systems
by locally manipulating gate voltages, tunneling barriers or magnetic
fields. This way, one may implement exact prescriptions to prepare those
long-lived collective excitations as many times as necessary to perform
specific tasks in a quantum computer \cite{preskill}. Those states (which
are not stationary states), are obtained by constructive interference from a
macroscopically dense number of quantum states. This can be seen from the
fact that the degeneracy of stationary points and the critical condition are
associated to nesting of the one-particle spectrum around inflection points
of the dispersion relation \cite{nest}. This peculiarity confers a degree of
\ `robustness' that is still present in finite systems, as shown in our
example with $N=100$, and leads one to expect that quantum coherence will be
maintained within operational levels when in contact with the \ `outside
world'.

\begin{center}
\textbf{ACKNOWLEDGMENTS}
\end{center}

One of the authors (DT) acknowledges support from brazilian CNPq through a
graduate scholarship. The other authors (JGC and GGC) are grateful to FAPESP
(S\~{a}o Paulo, Brazil) and CAPES (Brazil) for partial financial support.
GGC would also like to acknowledges support from FAEP (UNICAMP, S\~{a}o
Paulo, Brazil) for visiting the International Centre for Theoretical Physics
in Trieste, where part of this work was written.

\begin{center}
\textbf{FIGURE CAPTIONS}
\end{center}

\textbf{Fig. 1}

The inverse of the relaxation times (adimensional units) associated with the
stationary points of $\Omega ^{+}$\ as a function of $\gamma $ for the value 
$Q=2\pi /5$. They were obtained through the asymptotic behavior at very long
time ($t\rightarrow \infty $) of the infinite chain. Here, the wave vector $%
Q $ is fixed and the anisotropy is varied, crossing through the critical
value $\gamma _{C}=0.528$, which is indicated by an arrow. At $\gamma _{C}$,
two branches are degenerate, and the corresponding relaxation times
asymptotically diverge. Note that the branch of $\tau _{3}$ does not exist
for $\gamma >\gamma _{C}$.

\bigskip

\textbf{Fig.2}

Typical time evolution of the magnetization $<S_{Q}^{z}>(t)$ in a finite
system, for a value of $\gamma $ away from criticality. The $Q$ wave vector
is the same of Fig.1, and the subindex $n$ means that the magnetization is
normalized to the initial value at $t=0$. A regular relaxation behavior is
obtained until $\omega _{J}t\approx 250$, after which we observed the onset
of the first partial reconstruction. The exponent of relation (\ref{powerlaw}%
) is fitted using data from the upper (squares) and lower (circles)
envelopes. In the displayed case, we obtained $\nu _{0}(upper)\approx 0.476$
and $\nu _{0}(lower)\approx 0.482$, which are close to the predicted value $%
\nu _{0}=1/2$ for the infinite chain.

\bigskip

\textbf{Fig. 3}

The exponent $\nu _{0}$ from numeric calculations for $N=100$ spins as a
function of the anisotropy parameter $\gamma $, for the same $Q$ wave vector
of the preceding figures. The analytic calculation for the infinite chain
predicts a critical value $\gamma _{C}=0.528$, where the exponent \ $\nu
_{0} $ \ jumps from $1/2$ to $1/4$. The error bar represents the dispersion
obtained from estimating the exponent using the upper and lower envelopes of
the relaxation (see Fig. 2).

\end{document}